\documentclass[showpacs,preprintnumbers,
amsmath,amssymb,groupedaddress,superscriptaddress]{revtex4}
\usepackage{graphicx}
\usepackage{dcolumn}
\usepackage{bm}

\begin{document}
\title{Single and low-lying states dominance in two-neutrino 
double-beta decay}

\author{O. Moreno}
\affiliation{Instituto de Estructura de la Materia, CSIC, Serrano 123, 
E-28006 Madrid, Spain}
\author{R. \'{A}lvarez-Rodr\'{i}guez}
\affiliation{Department of Physics and Astronomy, University of Aarhus, 
DK-8000 Aarhus C, Denmark}
\author{P. Sarriguren}
\affiliation{Instituto de Estructura de la Materia, CSIC, Serrano 123, 
E-28006 Madrid, Spain}
\author{E. Moya de Guerra}
\affiliation{Departamento de F\'isica At\'omica, Molecular y Nuclear, 
Universidad Complutense de Madrid, E-28040 Madrid, Spain}
\author{F. \v Simkovic}
\affiliation{Department of Nuclear Physics, Comenius University, 
SK-842 15 Bratislava, Slovakia}
\author{A. Faessler}
\affiliation{Institut f\"ur Theoretische Physik, Universit\"at T\"ubingen, 
D-72076 T\"ubingen, Germany}

\date{\today}

\begin{abstract}

A theoretical analysis of the single-state dominance hypothesis
for the two-neutrino double-beta decay rates is performed on the 
examples of the double-beta decays of $^{100}$Mo, $^{116}$Cd, and 
$^{128}$Te. We also test the validity of an extended low-lying-state 
dominance that takes into account the contributions of the low-lying 
excited states in the intermediate nucleus to the double-beta decay 
rates. This study has been accomplished for all the double-beta 
emitters for which we have experimental information on their 
half-lives. The theoretical framework is a proton-neutron 
quasiparticle random-phase approximation based on a deformed Skyrme 
Hartree-Fock mean field with pairing correlations. Our calculations 
indicate that there are not clear evidences for single- or 
low-lying-state dominance in the two-neutrino double-beta decay.
Finally, we investigate the single electron energy distributions 
of the outgoing electrons in the double-beta decay processes with
an exact treatment of the energy denominators, which could help to
a more comprehensive analysis of NEMO-3 data.
\end{abstract}

\pacs{21.60.Jz;23.40.Hc;27.60.+j}

\maketitle

\section{Introduction}

The nonzero mass of the neutrino has been recently confirmed by neutrino 
oscillation experiments \cite{osci}, where neutrinos created in a flavor 
eigenstate are subsequently found to be in various flavors. However, 
these experiments can only probe differences of squared neutrino masses, 
but not the absolute neutrino mass scale, which remains still unknown. 
The neutrinoless double-beta ($0\nu\beta\beta$) decay nuclear process 
\cite{2bb1,2bb2,2bb3} is considered to be one of the most suitable 
candidates to provide this information. $0\nu\beta\beta$ decay is a 
lepton-number violating mode, which is forbidden in the Standard Model 
and that only occurs if neutrinos are massive Majorana particles. The 
half-life of this process involves an effective neutrino mass that could 
be eventually extracted from the measured half-lives.

Since the double-beta decay process is necessarily dependent on the nuclear
structure properties, lack of accuracy in the determination of the nuclear 
matrix elements involved in the process is a source of uncertainty in the 
information on neutrino properties that can be extracted from 
$0\nu\beta\beta$ decay experiments. Contrary to the $0\nu\beta\beta$ decay, 
the double-beta decay with the emission of two (anti)neutrinos 
($2\nu\beta\beta$) can proceed as a perturbative process in the Standard 
Model and has indeed been observed in several nuclides. This Standard 
Model allowed form of the decay is used as a test for the nuclear models 
considered. The operators involved in the $0\nu$ and $2\nu$ modes are 
different. In the $2\nu$ decay mode the operator connects the initial 
and final $0^+$ states via virtual $1^+$ transitions to the intermediate 
nucleus. In the $0\nu$ decay mode the transitions to the intermediate 
nucleus may take place to many different multipolarities. Nevertheless,
the underlying nuclear structure involved in both processes is similar. 
Success in describing the $2\nu$ decay mode is a requirement for a reliable 
calculation of the nuclear matrix elements related to the $0\nu\beta\beta$ 
decay.

The nuclear structure calculation involved in the $2\nu\beta\beta$ decay is 
not an easy task. Being a second order process in the weak interaction, the 
nuclear matrix elements involve a summation over the full set of virtual 
nuclear $1^+$ states in the odd-odd intermediate nucleus. Two leading 
microscopic nuclear models are commonly used to evaluate the nuclear matrix 
elements involved in the decay process. They are the nuclear shell model 
\cite{sm1,sm2} and the proton-neutron quasiparticle random-phase 
approximation (pnQRPA)\cite{pnqrpa1,pnqrpa2,pnqrpa3,pnqrpa4}. The shell 
model approach takes into account all possible correlations but within a 
restricted valence space. It successfully describes the low-lying excited 
states, but has difficulties with the description of states at high 
excitation energies, in particular, in the region of the Gamow-Teller (GT) 
resonance for open-shell medium and heavy nuclei, where most of the 
double-beta emitters are located. On the other hand, pnQRPA calculations 
do not have these problems related to the limited model space, but 
relevant ground state correlations might be missing. In addition, these 
calculations are very sensitive to model parameters. One problem of the 
theoretical $2\nu\beta\beta$ decay studies is whether the contributions 
of higher-lying states to the $2\nu\beta\beta$ decay amplitude, which 
are apparently disfavored by large energy denominators, play an important 
role. As we have mentioned, this has important consequences on the 
reliability of the shell model and pnQRPA to describe the nuclear matrix 
elements.

In the eighties, Abad and collaborators suggested \cite{aba84} that in 
those cases where the ground state of the odd-odd intermediate nucleus is 
a 1$^+$ state reachable by a GT transition, the transition through it
could account for the major part of the $2\nu\beta\beta$ matrix element. 
This conjecture is called the single-state dominance (SSD) hypothesis.
If SSD hypothesis is confirmed, the half-lives for $2\nu\beta\beta$ decay 
could be determined from single $\beta^-$ and electron capture (EC) 
measurements, experiments which are in principle much easier to carry out 
than the elaborated and time-consuming double-beta decay experiments.
On the theoretical side, the possible realization of the SSD hypothesis 
for the ground-state to ground-state transitions would lead to a drastic
simplification in the theoretical description of the intermediate nucleus,
since now only the lowest $1^+$ wave function has to be calculated. If
SSD is valid, this would increase the reliability of the small valence 
spaces used in the full shell model calculations.

From the experimental point of view, early tests of the SSD hypothesis
were evaluated on the $2\nu\beta\beta$ decay of $^{100}$Mo and $^{116}$Cd, 
two cases where the ground state of the intermediate odd-odd nucleus is
$J^\pi =1^+$. In Refs.~\cite{garcia} and \cite{bhatta}, the EC half-lives
of the ground state of the intermediate nucleus, $^{100}$Tc and $^{116}$In
respectively, were measured. Combining these measurements with the already 
known $\beta ^-$ half-lives, the SSD hypothesis was confirmed to some extent
on those 
examples. Experimental information on the EC and $\beta ^-$ decay of 
$^{128}$I is also available \cite{kanbe01}.

Theoretically, the SSD hypothesis was systematically studied in 
Ref.~\cite{civi_suho_98} within a pnQRPA formalism. The analysis performed 
there showed that the SSD hypothesis is realized in most cases but the 
mechanism leading to this property is not unique. In some instances the 
SSD is realized through a true dominance of the first $1^+$ virtual state, 
while in other cases it is realized due to cancellations among the 
contributions from higher lying $1^+$ states of the intermediate nucleus.
However, one should notice that the theoretical formalism used in 
Ref.~\cite{civi_suho_98}, is complemented with some adjusting of the 
single-particle levels around the proton and neutron Fermi surfaces to 
reproduce the observed single-quasiparticle spectrum. In addition, the 
scaling factor of the strength in the particle-particle residual force is 
fixed in each nucleus by optimizing the agreement with the experimentally 
known $EC$ and $\beta^-$ decay rates. This indicates that the theoretical 
evidences of the SSD hypothesis claimed in Ref.~\cite{civi_suho_98} rely 
on the experimental ones.

In Ref.~\cite{simk_ssd} the SSD hypothesis was studied considering an 
exact treatment of the energy denominators of the perturbation theory. It 
was found that the $2\nu\beta\beta$ half-lives are reduced by a factor of 
about 20\% for ground-state to ground-state transitions. It was also shown 
that by measuring the single-electron spectra and/or the angular 
distributions of the emitted electrons, the SSD hypothesis could be 
confirmed or ruled out experimentally in the near future.

In the last years, experimental information on the low-lying Gamow-Teller 
strength coming from charge-exchange reactions has been collected for
several nuclei relevant to double-beta decay processes 
\cite{frekers,madey,aki97,sas07,rak05}. In particular, 
($p,n$) reactions were used in Ref.~\cite{madey} to study the transitions 
$^{128}$Te$(0^+)$ $\to$ $^{128}$I$(1^+)$, the ($^3$He,$t$) reaction was 
used in Ref.~\cite{aki97} to study the transitions $^{100}$Mo$(0^+)$ 
$\to$ $^{100}$Tc$(1^+)$ and $^{116}$Cd$(0^+)$ $\to$ $^{116}$In$(1^+)$. 
The ($p,n$) reaction was also used in Ref.~\cite{sas07} to study
$^{116}$Cd$(0^+)$ $\to$ $^{116}$In$(1^+)$. Similarly, the reaction 
($d$,$^2$He) was used in Ref.~\cite{rak05} to study the transitions
$^{116}$Sn$(0^+)$ $\to$ $^{116}$In$(1^+)$. These measurements complement 
the experimental information already available on the $ft$ values for 
the transitions from the $1^+$ ground state of the odd-odd intermediate 
nucleus to the $0^+$ ground states of parent and daughter nuclei. In 
addition, they give information on the transitions to the $1^+$ excited 
states of the intermediate nucleus.

In this paper we use a formalism based on a pnQRPA approach with a 
selfconsistent quasiparticle basis obtained from deformed Skyrme Hartre-Fock 
calculations with pairing correlations \cite{vautherin}. On top of that we 
include separable residual spin-isospin interactions in both particle-hole 
and particle-particle channels \cite{single}. Within this formalism we 
calculate the GT strength distributions of the single beta branches as 
well as the $2\nu\beta\beta$ matrix elements corresponding to the 
double-beta emitters $^{100}$Mo, $^{116}$Cd, and $^{128}$Te. With these 
results, we first evaluate to what extent the SSD hypothesis or the more 
relaxed low-lying states dominance (LLSD) hypothesis are fulfilled within our
theoretical approach. Comparison with the available experimental information
on both single-beta branches, $EC/\beta^-$, and  $2\nu\beta\beta$ decay is 
also made. Finally, we extend the LLSD analysis to the rest of the observed 
$2\nu\beta\beta$ emitters, where the ground state of the  intermediate 
nucleus is not $1^+$, and we evaluate the contributions to the 
$2\nu\beta\beta$ half-lives considering an increasing range of excitation 
energy in the intermediate nucleus. Former calculations of running sums 
of matrix elements as a function of the excitation energy of the $1^+$  
intermediate nucleus were presented in Refs.~\cite{simk_def,nakada}.
In Ref.~\cite{simk_def} the results presented were for $^{76}$Ge and were
obtained within a phenomenological Woods-Saxon mean-field approach.
In Ref.~\cite{nakada} results were presented for $^{36}$Ar, $^{54}$Fe,
and $^{58}$Ni within a shell model approach.

The paper is organized as follows. In Sec. II we present briefly the main
formalism and discuss various approximations. In Sec. III we compare our
results with experiment in the cases of $^{100}$Mo, $^{116}$Cd, and 
$^{128}$Te and analyze the possible realizations of SSD and LLSD hypotheses. 
We also study in this section the single electron energy distributions 
of the outgoing electrons in the $2\nu\beta\beta$ decay processes with an
exact treatment of the energy denominators.
In  Sec. IV we study the contributions of the different energy 
ranges to the $2\nu\beta\beta$ matrix elements in all the observed 
$2\nu\beta\beta$ emitters and discuss them in terms of LLSD. Finally we 
present in Sec. V the summary and the main conclusions.

\section{Brief description of theoretical calculations}

The double-beta decay is a nuclear process characterized by a change of two 
units in the nuclear charge, while the atomic mass number remains unchanged. Here 
we focus on the $2\nu\beta^-\beta^-$ decay mode where two antineutrinos 
and two electrons are emitted,

\begin{equation}
(Z,A)_{0^+_i}  \to (Z+2,A)_{0^+_f} + 2e^- + 2\bar{\nu}_e \, .
\end{equation}

Only the basic expressions are written here, more details can be found 
in Refs.~\cite{simk_def,raquel}. The half-life of the $2\nu\beta\beta$ decay 
can be written as 

\begin{equation}\label{half-life}
\left[ T_{1/2}^{2\nu\beta\beta}\left( 0^+_{\rm gs} \to 0^+_{\rm gs}  
\right) \right] ^{-1}= G^{2\nu\beta\beta} 
\left| M_{GT}^{2\nu\beta\beta}\right| ^2 \, ,
\end{equation}
where $G^{2\nu\beta\beta}$ is the phase-space integral \cite{2bb1,sm1}.
The nuclear matrix element $M_{GT}^{2\nu\beta\beta}$ contains all the 
information on the nuclear structure involved in the process. The 
theoretical formalism used here is the pnQRPA based on a selfconsistent 
mean field that allows for pairing and deformation. We refer to this 
formalism in short as deformed HF+BCS+QRPA. In this formalism the nuclear 
matrix element can be written as,

\begin{equation}\label{mgt1}
M_{GT}^{2\nu\beta\beta}=\sum_M\sum_{\mu_i,\mu_f} (-1)^M \frac{\langle 0^+_ f 0| 
\sigma_{-M}t^- | 1^+_{\mu_f}M \rangle \:\langle 1^+_{\mu_f}M | 1^+_{\mu_i}M 
\rangle \: \langle 1^+_{\mu_i}M | \sigma_Mt^- | 0^+_i 0\rangle }
{(\omega^{\mu_f}+ \omega^{\mu_i}) / 2} \, ,
\end{equation}
where the intermediate states are the $1^+$ excitations of the initial 
and final nuclei in the laboratory frame. We use the Bohr-Mottelson
factorization \cite{bmot}
for the total angular momentum states $I^+$ in the laboratory frame.
The index $\mu_i (\mu_f)$ in Eq. (\ref{mgt1}) contains the $K_i(K_f)$
third component of the angular momentum
of the QRPA state in the intrinsic system,

 \begin{equation}\label{bohrmot}
| I^+_{\mu_i} M \rangle \equiv | I^+_{m_i} K_iM \rangle = \sqrt 
{\frac{2I+1}{16\pi^2 (1+\delta_{K_i,0})}}
\left[ D^{\dagger \,I}_{K_iM} (\Omega) | \phi ^{m_i}_{K_i} \rangle +
(-1)^{I-K_i} D^{\dagger \,I}_{-K_iM} (\Omega) | \bar \phi ^{m_i}_{K_i} \rangle 
\right] \; ,
\end{equation}
with the intrinsic state

\begin{equation}
|\phi^{m_i}_{K_i}\rangle = \Gamma ^{m_i}_{K_i}|\phi_{0,i}\rangle\; 
\qquad K_i=0,\pm 1,
\end{equation}
with $|\phi_{0,i}\rangle $ the QRPA vacuum of the initial state and 
$\Gamma ^{m_i}_{K_i}$ the phonon QRPA operators.
$| \bar\phi ^{m_i}_{K_i} \rangle$ are the time reversed of  
$| \phi ^{m_i}_{K_i} \rangle$.
Similarly, the $0^+$ initial and final states are of the Bohr-Mottelson form

\begin{equation}\label{bohr0}
| 0^+_{i} 0 \rangle = \frac{1}{\sqrt{8\pi^2}} D^{\dagger \, 0}_{00} (\Omega) 
| \phi_{0,i} \rangle \; ,
\end{equation}
and similarly for the final state $f$.
Integrating over the angular $\Omega$ variables, we finally obtain

\begin{equation}\label{mgt}
M_{GT}^{2\nu\beta\beta}=\sum_{K=0,\pm 1}\sum_{m_i,m_f} (-1)^K \frac{\langle 
\phi_{0,f}| \sigma_{-K}t^-| \phi^{m_f}_K \rangle \:\langle \phi^{m_f}_K | 
\phi^{m_i}_K \rangle \: \langle \phi^{m_i}_K  | \sigma_Kt^- | \phi_{0,i} \rangle}
{(\omega_K^{m_f}+ \omega_K^{m_i}) / 2} \, .
\end{equation}
As it is well known (see Ref. \cite{emoya} and references therein) the 
Bohr-Mottelson factorization
of the wave function gives the adiabatic limit of the exact angular momentum
projection and is a very good approximation for well deformed nuclei. First
order corrections to the Bohr-Mottelson factorization due to angular momentum
projection are typically less than $1/\langle J^2_{\perp}\rangle $, where 
$\langle J^2_{\perp}\rangle $ is larger than 100 for well deformed nuclei
\cite{berdi}.
In Eq. (\ref{mgt}) $\omega_K^{m_i} (\omega_K^{m_f})$ are the RPA excitation energies 
of the intermediate $1^+$ virtual states with respect to the initial 
(final) nucleus, with energy $E_i (E_f)$. The energy denominator is the 
average excitation energy of a couple of these intermediate states. The 
indices $m_i$, $m_f$ run over all the 1$^+$ states of the intermediate 
nucleus ($Z+1,A$) in the transition, and $K=0,\pm1$ are the possible 
angular momentum projections of these intermediate states. The overlaps 
are needed to take into account the nonorthogonality of the intermediate 
states reached from initial and final ground states. Their expressions 
can be found in Refs.~\cite{simk_def,raquel}.

To compute the matrix element we proceed as follows. Each initial and
final even-even ground state is described as a pnQRPA vacuum. For 
each $(N,Z)$ nucleus, a BCS ground state is independently obtained from a 
selfconsistent mean-field calculation on a large axially-symmetric harmonic 
oscillator basis, using a density-dependent effective interaction and a 
phenomenological pairing-gap parameter. The nucleon-nucleon effective 
force  SLy4 \cite{cha98} is used to generate the HF mean field, and the 
BCS equations are solved at each iteration with a value of the 
pairing gap  obtained from even-odd experimental mass differences
between neighboring nuclei \cite{au03}. When the iterative process
converges, a set of single-particle states is obtained with their 
corresponding wave functions, energies, and occupation probabilities. 
It is worth noticing that the quadrupole deformation of the nucleus
in its ground state is obtained selfconsistently as the shape that
minimizes the nuclear energy. Separable Gamow-Teller particle-hole 
($ph$) and particle-particle ($pp$) residual interactions are included 
and treated in pnQRPA \cite{single}. The corresponding coupling constants 
$\chi_{ph}$ and $\kappa_{pp}$ have been used in this work with values  
$\chi_{ph}=20/A$ MeV and $\kappa_{pp}=3/A$ MeV, which give a good 
description of GT properties when SLy4 Skyrme force is used, as we 
shall show in next section.

Due to the many-body approximations performed in the calculation, the 
intermediate 1$^+$ states computed from the initial and from the final 
ground states are different, and a projection from one basis to the other 
is in order. To get the $2\nu\beta\beta$ decay matrix element, every 
possible transition through all the intermediate 1$^+$ states is taken 
into account. The amplitude of each transition is divided by an energy 
denominator, as shown in Eq. (\ref{mgt}). The energy denominators written 
in this way are obtained by replacing the lepton energies 
$E_{ei} +E_{\nu j}$, where $E_e$ ($E_\nu$) is the energy of the 
emitted electron (antineutrino), by an average quantity \cite{2bb1,2bb2,2bb3}

\begin{equation}
E_{ei} + E_{\nu j} \approx \frac{1}{2}Q_{\beta\beta}+m_e=\frac{1}{2}
(E_i-E_f) \, ,
\end{equation}
where $Q_{\beta \beta}$ is the $Q$-value of the double-beta decay 
$(Q_{\beta \beta}=E_i -E_f -2m_e)$. $E_i$ and $E_f$ are the ground-state 
energies of parent and daughter nuclei, respectively. The consequences of 
this approximation on the SSD tests have been studied in
Ref.~\cite{simk_ssd}. We shall come back to this point in the next section. 
Different ways of writing these energy denominators, which are all equivalent, 
can be found in the literature. In particular one finds

\begin{equation}
\frac{1}{2}(\omega_K^{m_f}+\omega_K^{m_i})= 
\frac{1}{2}Q_{\beta\beta}+E_m-E_i+m_e =
E_m-\frac{1}{2}\left( E_i+E_f \right)  \, ,
\end{equation}
where $E_m$ is the energy of the intermediate nucleus in the state $m$. 
In the cases where the ground state of the intermediate nucleus is a $1^+$
state, the denominator corresponding to the decay passing through this state
can also be written as $( Q_{\beta^-} + Q_{\rm EC})/2$, where $Q_{\beta^-}$ 
and $Q_{\rm EC}$ are the $Q-$values corresponding to its $\beta ^-$ and EC 
decay, respectively.

In this paper we study the contributions to the $2\nu\beta\beta$ nuclear 
matrix elements from consecutive terms in Eq. (\ref{mgt}) with increasing 
energy denominators. The main purpose is to check 
the validity of the SSD hypothesis by considering only the lowest-energy 
term in the summation in the cases of $^{100}$Mo, $^{116}$Cd, and $^{128}$Te, 
which are emitters with $1^+$ intermediate ground states. However, it is also 
interesting to check the extent to which the contributions coming from 
low-lying virtual excitations are able to reproduce the total $2\nu\beta\beta$
nuclear matrix elements in a more general way. 
Actually, the strong fragmentation of the GT strength within the deformed
formalism suggests that a meaningful comparison of theoretical results with 
experiment should be done for the accumulated strength in a given energy range
rather than a direct comparison of individual excitations. Thus, we analyze 
the summation in Eq. (\ref{mgt}) considering only the contribution of the 
intermediate ground state (to test SSD), then the contribution corresponding 
to all the low-lying states (to test LLSD), and finally taking into account 
all possible contributions (to get the complete result).

As mentioned before, in the cases where the ground state of the intermediate 
nucleus is a $1^+$ state, we have experimental information on the 
$\beta^-$ and EC decays of the ground state intermediate nucleus into the 
$0^+$ ground states of the daughter and parent nuclei, respectively. We also 
have complementary experimental information based on charge-exchange reactions 
that can be evaluated and compared to the $2\nu\beta\beta$ experimental nuclear 
matrix element, as well as to the theoretical calculations. The contribution 
involving low-lying states can also be evaluated from the experimental 
information including now charge-exchange reactions \cite{aki97,rak05}, and 
can be compared again with the total $2\nu\beta\beta$ experimental nuclear 
matrix element, as well as to the theoretical calculations.

\section{Results for $2\nu\beta\beta$ decays with $1^+$ ground state 
intermediate nuclei}

\subsection{Single-state dominance}

We start this section by summarizing the experimental situation concerning the 
validity of the SSD hypothesis in the $2\nu\beta\beta$ decay of $^{100}$Mo, 
$^{116}$Cd, and $^{128}$Te. The analysis is based on information extracted from 
ground-state to ground-state decay measurements of the intermediate nuclei 
$^{100}$Tc, $^{116}$In, and $^{128}$I, as well as on information extracted from 
charge-exchange reactions. Table I, II, and III contain the GT matrix elements 
for ground state to ground state transitions in the cases $A=100$, $A=116$, 
and $A=128$, respectively. The notation used corresponds to the reaction point 
of view, where the $B(GT^-)$ strength corresponds to the $(Z,A)\to (Z+1,A)$ 
transition and the $B(GT^+)$ strength corresponds to the $(Z+2,A)\to (Z+1,A)$ 
transition. Thus, in the case of the decay of the intermediate nucleus, we 
still call $B(GT^-)$ to the strength corresponding to the $EC/\beta^+$ decay, 
and  $B(GT^+)$ to the strength corresponding to the $\beta^-$ decay,
according to the following scheme:

\begin{center}
\begin{tabular}{|lclcl|}
\hline
&&&& \\
& $B(GT^-)$ && $B(GT^+)$ & \\  
&&&&\\
\hline
&&&& \\
& $\stackrel{\textstyle EC}{\longleftarrow}$ && $\stackrel{\textstyle \beta^-}
{\longrightarrow}$ & \\
& $\stackrel{\textstyle \longrightarrow}{(^3{\rm He},t)}$ &&
$\stackrel{\textstyle \longleftarrow}
{(d,^2{\rm He})}$ & \\
& $(p,n)$ && $(n,p)$ & \\
$^{100}$Mo && $^{100}$Tc && $^{100}$Ru \\
$^{116}$Cd && $^{116}$In && $^{116}$Sn \\
$^{128}$Te && $^{128}$I && $^{128}$Xe \\
&&&& \\
\hline
\end{tabular} 
\end{center}
In this paper, results for $B(GT)$ are given in units in which the neutron
decay has $B(GT)$=3. The matrix elements are extracted from the cross 
sections in the case of charge-exchange reactions \cite{rak05} and from the 
$\log ft$ values in the case of the $\beta$ decay :

\begin{equation}
M(GT)=\left[ \frac{3D}{g_A^2\, ft}\right] ^{1/2} \, ,
\end{equation}
where $D=6147$ and $g_A=1.25$ is the axial-vector coupling strength.
The GT matrix element for the $2\nu\beta\beta$ decay within the SSD
hypothesis is obtained from 

\begin{equation}
M_{GT}^{2\nu\beta\beta}(SSD)=\frac{M(GT^-)M(GT^+)}{(Q_{\beta^-} + 
Q_{\rm EC})/2}= \frac{1}{\sqrt{ft_{\rm EC}}}\frac{1}
{\sqrt{ft_{\beta^-}}}\frac{6D}{g_A^2(Q_{\beta^-} + Q_{\rm EC})}\, , 
\label{dgt}
\end{equation}
where the overlap in Eq. (\ref{mgt}) has been approximated by one.

Table I shows the experimental Gamow-Teller matrix elements $M(GT^-)$ 
and $M(GT^+)$ for the transitions connecting the $1^+$ ground state in 
$^{100}$Tc with the $0^+$ ground states in $^{100}$Mo and $^{100}$Ru. 
The $M(GT^-)$ matrix elements were extracted from ($^3$He,$t$) 
charge-exchange reactions \cite{aki97} and from the $ft_{\rm EC}$ value 
as given in  Ref.~\cite{garcia}. The $M(GT^+)$ matrix element was obtained 
from the $ft_{\beta^-}$ value \cite{sin08} in the decay 
$^{100}$Tc $\to$ $^{100}$Ru. The matrix elements $M_{GT}^{2\nu\beta\beta}$ 
are obtained from the two possible combinations of data. The
half-lives $T_{1/2}^{2\nu\beta\beta}$ are calculated using 
approximated (SSD1) and exact (SSD2) energy denominators (we keep the 
notation used in previous references \cite{simk_ssd}).
The experimental half-life is 
$T_{1/2}^{2\nu\beta\beta}=7.1\times 10^{18}$ y \cite{bar0206}, 
which gives rise to a matrix element
$(M^{2\nu\beta\beta}_{GT})_{\rm exp}= 0.241$ MeV$^{-1}$, using $g_A=1.25$.
One should notice that the half-lives are independent on the $g_A$ value.
As one can see in Table I, the agreement with the experimental half-life
improves with SSD2 approach and it is particularly good when the matrix
elements extracted from  $\log ft$ are used.

Table II is similar to Table I but for $A$=116.
Experimental values for $M(GT^-)$ are taken from ($^3$He,$t$)
\cite{aki97} and ($p,n$) \cite{sas07} charge-exchange reactions, and from 
the $ft_{\rm EC}$ value for the decay of $^{116}$In into the ground state 
of $^{116}$Cd reported in Ref.~\cite{bhatta}. Experimental values for 
$M(GT^+)$ are from Ref.~\cite{rak05}, where the GT strength was extracted 
from the ($d$,$^2$He) charge-exchange reaction $^{116}$Sn $\to$ $^{116}$In. 
In this reference, the strength was normalized to recover at zero excitation 
energy the $\beta^-$ decay data \cite{bla01}. The results for 
$M_{GT}^{2\nu\beta\beta}$ are again compared to the values extracted from 
the measured $2\nu\beta\beta$ decay half-life \cite{bar0206} 
($T_{1/2}^{2\nu\beta\beta}=3.0\times 10^{19}$ y,  and a corresponding 
matrix element
$(M^{2\nu\beta\beta}_{GT})_{\rm exp}= 0.127$ MeV$^{-1}$, using $g_A=1.25$).
Except for the first row that contains the values extracted from
($^3$He,$t$) and that clearly overestimates the experimental half-life
by almost one order of magnitude, we observe that both SSD1 and SSD2 
half-lives underestimate the measured value.

Table III contains similar information for the case $A=128$. Experimental 
values for $M(GT^-)$ are taken from the $ft_{\rm EC}$ value for the decay of 
$^{128}$I into the ground state of $^{128}$Te reported in  Ref.~\cite{kanbe01}, 
as well as from ($p,n$) reactions \cite{madey}. Experimental values for 
$M(GT^+)$ are taken from the $ft_{\beta^-}$ value for the decay of $^{128}$I 
into the ground state of $^{128}$Xe \cite{kanbe01}.
In this case both SSD1 and SSD2 predictions overestimate the experimental 
half-life \cite{bar0206}
($T_{1/2}^{2\nu\beta\beta}=2.5\times 10^{24}$ y, and a corresponding matrix 
element $(M^{2\nu\beta\beta}_{GT})_{\rm exp}= 0.043$ MeV$^{-1}$, using $g_A=1.25$). 

\subsection{Low-lying states dominance}

\begin{figure*}
\centering \includegraphics[width=130mm]{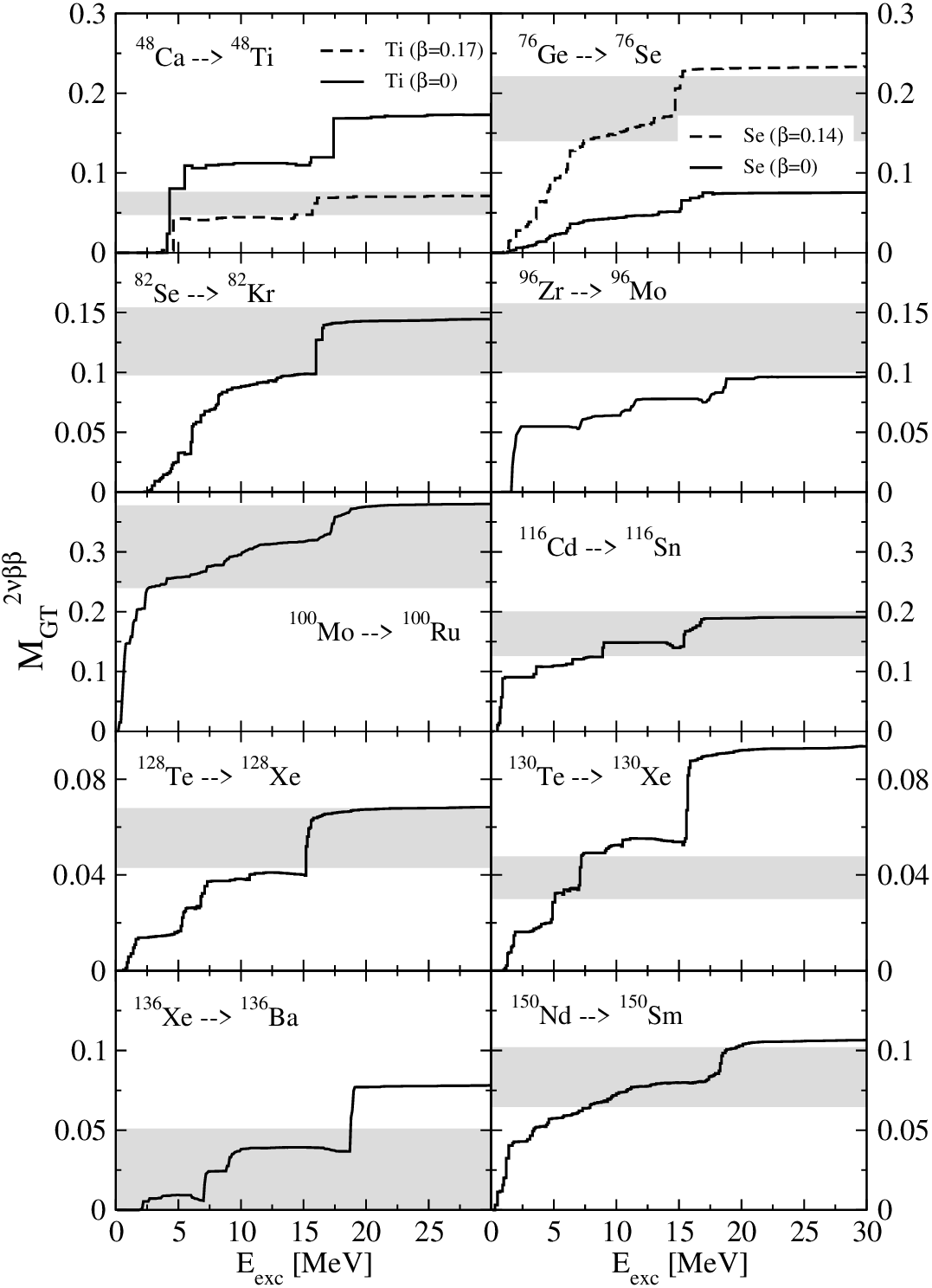}
\caption[]{Double-beta decay matrix elements as a function
of the intermediate nucleus excitation energy considered
in the calculation. An experimental range is given by 
a shadow area where the upper and lower limits are deduced
from the experimental half-lives \cite{bar0206}, using
$g_A=$1.00 and $g_A=$1.25, respectively. Dashed lines in $A=48$ and
$A=76$ correspond to calculations with an alternative choice
of deformation (see text).}
\label{bb_accum}
\end{figure*}

In Tables IV, V, and VI, for the cases $A=100$, $A=116$, and $A=128$ 
respectively, we show theoretical results and experimental data corresponding 
to the $B(GT^\pm$) strengths to the low-lying virtual states of the
corresponding intermediate nuclei, as well as the contributions to 
$2\nu\beta\beta$ decay matrix element of the virtual transitions through 
them. A quenching factor $g_{A,{\rm eff}}=0.8\, g_{A,{\rm free}}$
has been introduced in all the theoretical calculations reported in this
paper, except for those shown later on in Fig.~\ref{bb_accum}. 
In Table IV we do not have information on the low-lying $B(GT^+)$ 
strengths from charge-exchange reactions, only the $\log ft$ value is 
available. Thus, we only compute theoretical values for the 
$2\nu\beta\beta$ decay matrix element. We can see that the theoretical 
$B(GT^-)$ strengths accumulated up to 1.4 and 2.6 MeV, agree with the 
measurements \cite{aki97} within a 80-90\%. On the other hand, the 
$2\nu\beta\beta$ decay matrix element calculated up to 2.6 MeV of excitation 
energy in the intermediate nucleus accounts only for 65\% of the experimental 
value.

In the case of $A=116$ we can combine the experimental information available
from ($^3$He,$t$) \cite{aki97} and from ($d$,$^2$He) \cite{rak05} to evaluate
the $2\nu\beta\beta$ decay matrix element $M_{GT}^{2\nu\beta\beta}$ up to 3 
MeV, and we can compare the obtained values with the total matrix element 
evaluated from 
the measured $2\nu\beta\beta$ half-life. Thus, we can test to what extent 
LLSD is valid in this case both theoretically and experimentally. First, we 
observe in Table V that the two single branches are reasonably reproduced 
by the calculations. In the case of $B(GT^+)$ the agreement is apparent. 
In the case of $B(GT^-)$ the comparison to experiment is not so meaningful 
because there is no agreement between experimental data from different
groups \cite{aki97,sas07}. In particular, a very different value has been 
reported for the transition to the ground state. If we consider the new 
revised values given in Ref.~\cite{sas07} (see Table II), the agreement with 
the calculations clearly improves.
The $2\nu\beta\beta$ decay matrix elements evaluated at increasing energies
show first that they approach to the experimental value obtained from the 
$2\nu\beta\beta$ half-life (0.127 MeV$^{-1}$). We also observe that the 
theoretical results increase with energy to yield about a half of the 
experimental value at 3 MeV. Agreement improves when using the 
value of Ref.~\cite{sas07}. Table VI contains the results for $A=128$.
As in the case of $A=100$, we do not have information on the low-lying 
$B(GT^+)$ strengths. Thus, we only compute theoretical values for the 
$2\nu\beta\beta$ decay matrix element. 

\begin{figure*}
\centering \includegraphics[width=150mm]{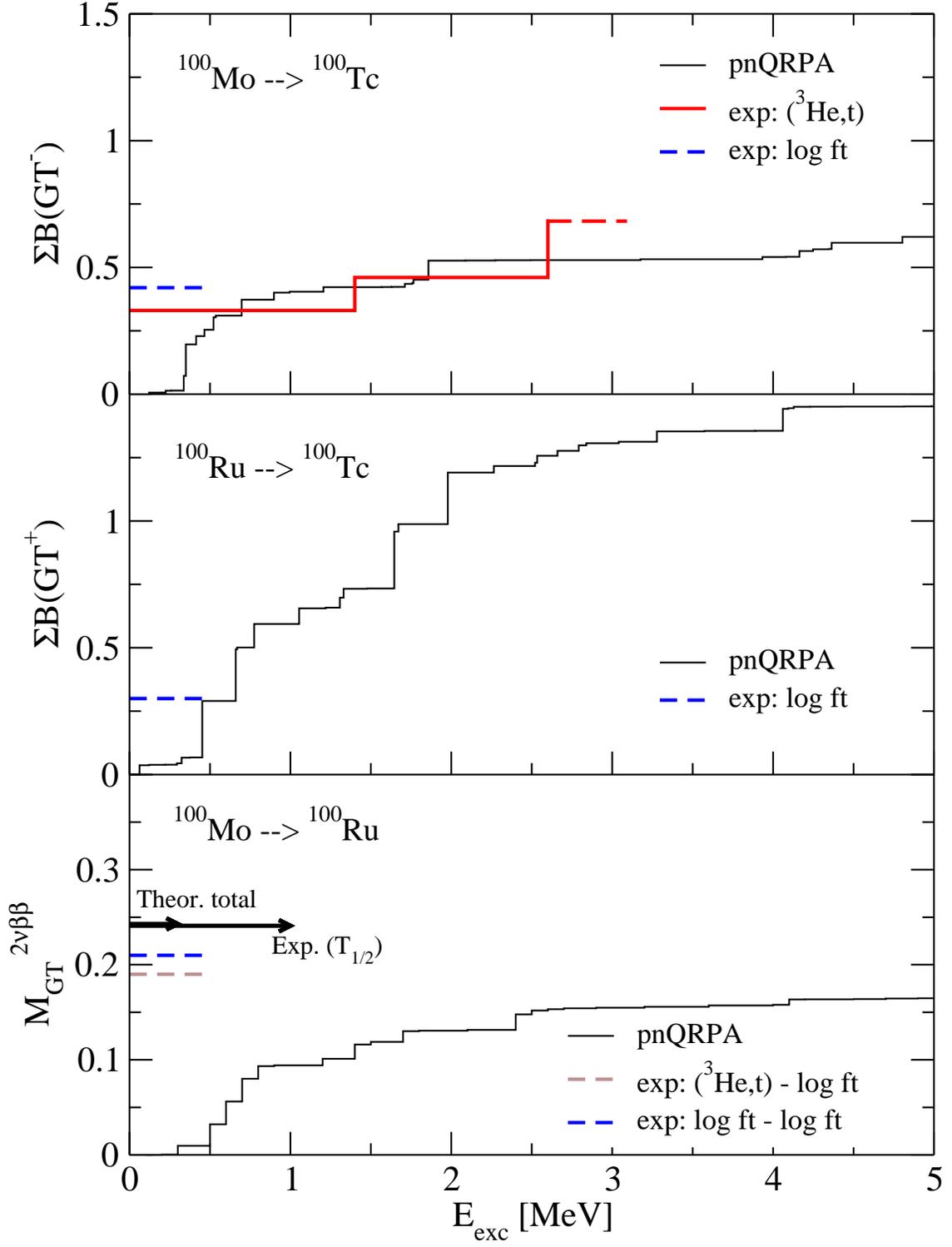}
\caption[]{Accumulated $B(GT^-)$ strength for the transition
$^{100}$Mo$\to^{100}$Tc (upper panel), accumulated $B(GT^+)$ strength
for the transition $^{100}$Ru$\to^{100}$Tc (middle panel) and 
$M^{2\nu\beta\beta}_{GT}$ [MeV$^{-1}$]
matrix element for the transition $^{100}$Mo$\to^{100}$Ru (lower
panel) as a function of the excitation energy of the intermediate
nucleus from a HF+BCS+QRPA calculations, together with experimental
data from various sources.  }
\label{100_b+_b-_bb}
\end{figure*}

Fig.~\ref{100_b+_b-_bb} shows the accumulated single $B(GT^-)$ strength
from $^{100}$Mo $\to$ $^{100}$Tc (upper panel), the accumulated single 
$B(GT^+)$ strength from $^{100}$Ru $\to$ $^{100}$Tc (middle panel) and the 
$2\nu\beta\beta$ GT matrix element from $^{100}$Mo $\to$ $^{100}$Ru (lower 
panel) in the low excitation energy range (up to 5 MeV). Available 
experimental data are shown by thick lines. 

As one can see in Fig.~\ref{100_b+_b-_bb} the theoretical $B(GT^-)$ strength
below 3 MeV is rather fragmented, but the total accumulated strength 
agrees well with experiment. Although not shown in the figure, it is worth 
noticing that the GT resonance measured in Ref.~\cite{aki97} at 13-14 MeV 
with a strength $B(GT^{-})=23$ is well reproduced in our calculation,
yielding a peak centered at 16 MeV with a similar strength. 
The strength obtained from the $\log ft$ value also agrees with the strength 
calculated up to about 0.5 MeV. In the case of the $B(GT^+)$ strength the only 
information comes from the $\log ft$ value at zero excitation energy. The 
corresponding strength is again well reproduced by the calculation of the 
accumulated strength below 0.5 MeV.

In the lower panel we show the theoretical calculation for the $2\nu\beta\beta$ 
matrix element as a function of the excitation energy taken into account in the 
sum in Eq. (\ref{mgt}). It is compared with the experimental matrix element 
extracted from the measured $2\nu\beta\beta$ half-life. We compare our results 
with the values extracted from the measured $\log ft$ and charge-exchange 
reactions to check the validity of the SSD hypothesis graphically. We also 
show the total theoretical value when all the intermediate $1^+$ states are 
considered in the calculation (see also Fig.~\ref{bb_accum}). As can be seen in 
the lower panel, a rapid increase in the $2\nu\beta\beta$ decay matrix element 
takes place up to 2 MeV of intermediate excitation energy. It accounts for 
about 60$\%$ of the total matrix element. Eventually (see Fig.~\ref{bb_accum}) 
the calculations reach the experiment when one considers the whole energy range.
From the single-beta strengths one can see that there are no especially strong 
transitions to any low-energy intermediate state (in particular, to the ground 
state), which prevents the $2\nu\beta\beta$ matrix element from showing the 
same effect. 

Summarizing, in the $2\nu\beta\beta$ decay of $^{100}$Mo, the SSD hypothesis
is fulfilled experimentally within 90\%.
The total theoretical $2\nu\beta\beta$ matrix element
agrees well with experiment, while LLSD calculations up to 
2 MeV account for 60\% of the experimental $2\nu\beta\beta$ matrix 
element. The remaining 40\% comes from contributions of states at higher
excitation energy.

\begin{figure*}
\centering \includegraphics[width=150mm]{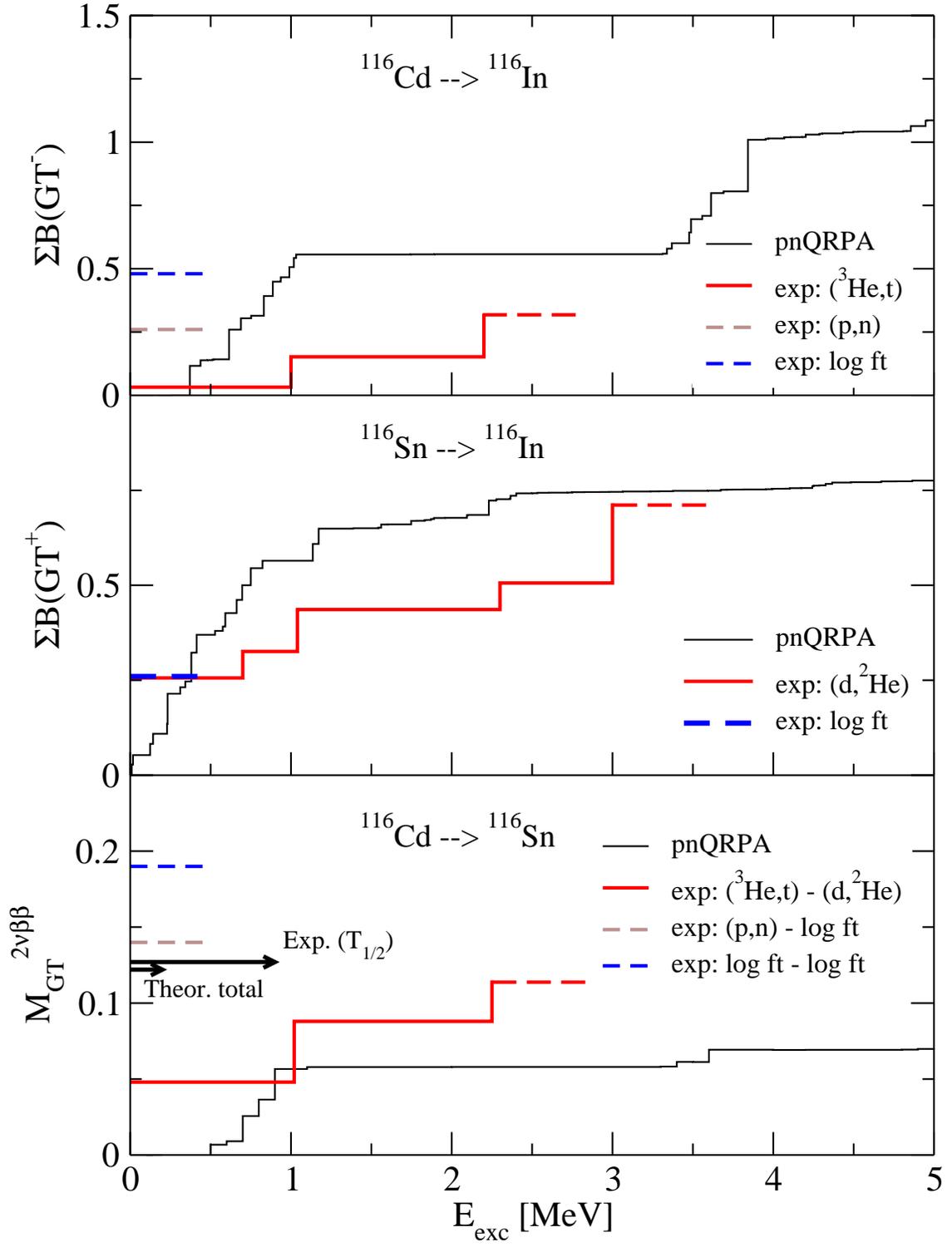}
\caption[]{Same as in Fig.~\ref{100_b+_b-_bb}, but for the 
transition  $^{116}$Cd$\to^{116}$Sn.}
\label{116_b+_b-_bb}
\end{figure*}

Same results are shown for the transitions $^{116}$Cd $\to$ $^{116}$In
(upper panel), $^{116}$Sn $\to$ $^{116}$In (middle panel) and $^{116}$Cd
$\to$ $^{116}$Sn (lower panel) in Fig.~\ref{116_b+_b-_bb}. In the upper 
panel we show experimental data for $B(GT^-)$ from ($^3$He,$t$), together 
with the data at zero excitation energy extracted from ($p,n$), and from 
the $\log ft$ values. We also show the theoretical calculations that produce
an increasing strength up to 1 MeV, which is very fragmented, but that 
finally reaches the most reliable experimental strength obtained from the
measured $\log ft$. In this case the GT resonance measured in 
Ref.~\cite{aki97} amounts to $B(GT^-)=26$ and it is located at 14.5 MeV.
Our calculation yields about the same strength centered at 16 MeV. 
The experimental data for $B(GT^+)$ are from ($d,^2$He) charge-exchange 
reactions \cite{rak05}, where the value extracted from $\beta^-$ decay was 
used to normalize the strength.

The $B(GT^+)$ strength in the middle panel shows that the calculation 
reproduces the total experimentally measured strength below 3 MeV. The
distribution of this strength is more fragmented in the calculations but
the general trend is similar. In the lower panel we compare again our 
theoretical results for the $2\nu\beta\beta$ decay matrix element with 
experiment. In this case we compare with  the experimental matrix element 
extracted from the measured half-life, as well as with the values 
extracted from the measured  $\log ft$ values, and with the values 
extracted from charge-exchange reactions.
We also show the total theoretical value (see also Fig.~\ref{bb_accum}), 
which is very close to the experiment.

Theoretically, as in the case of $A=100$, a strict SSD or LLSD is not 
observed in the lower panel. The few contributions below 1 MeV of excitation 
energy account for a 50$\%$ of the final value of the double-beta decay 
matrix element. Experimental accumulated single strengths are roughly in 
agreement with the results shown, but the double-beta experimental data for 
the low-energy contributions are, as in the case of $A=100$, larger than our
theoretical results.

Summarizing, in the $2\nu\beta\beta$ decay of $^{116}$Cd there is no clear
evidence that the SSD hypothesis is fulfilled experimentally 
because of the spread of data
coming from different sources. The experimental LLSD up to 2.5 MeV accounts
for the experimental $2\nu\beta\beta$ decay matrix element, while the 
calculations up to this energy account only for half of it. Since the total 
calculated matrix
elements agrees well with experiment, the remaining contribution comes
again from excited states at higher energies.

\begin{figure*}
\centering \includegraphics[width=150mm]{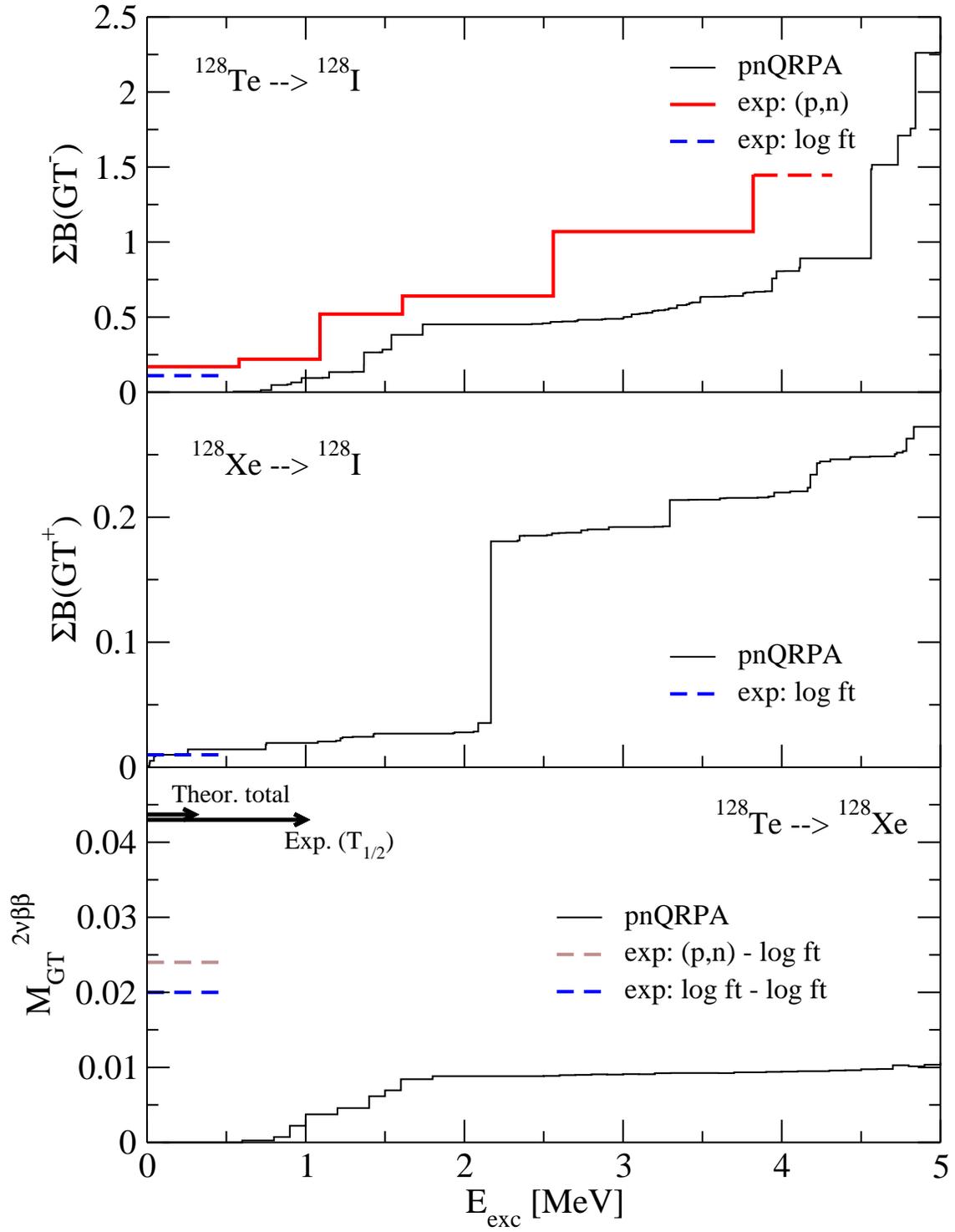}
\caption[]{Same as in Fig.~\ref{100_b+_b-_bb}, but for the 
transition  $^{128}$Te$\to^{128}$Xe.}
\label{128_b+_b-_bb}
\end{figure*}

Fig.~\ref{128_b+_b-_bb} is the corresponding figure for $A=128$.
In the upper panel we show experimental data for $B(GT^-)$ from ($p,n$), 
together with the data at zero excitation energy extracted from
the $\log ft$ values. We also show the theoretical pnQRPA calculations.
In this case the GT resonance measured in Ref.~\cite{madey} 
contains a strength $B(GT^-)=34.24$ located at 13.14 MeV. 
Our calculation yields $B(GT^-)=35$ centered at 15 MeV. 
In the middle panel the experimental value at zero energy is obtained
from $\log ft$ \cite{kanbe01} and agrees well with the calculation
(note the different scale as compared to the upper panel).
The lower panel contains the results for the $2\nu\beta\beta$ decay 
matrix element. The meaning is the same as in the previous figures.
We see that the total theoretical value (see also Fig.~\ref{bb_accum}), 
is very close to the experiment and it is not reproduced by 
SSD or LLSD hypotheses.

\begin{figure*}
\centering \includegraphics[width=60mm,angle=90]{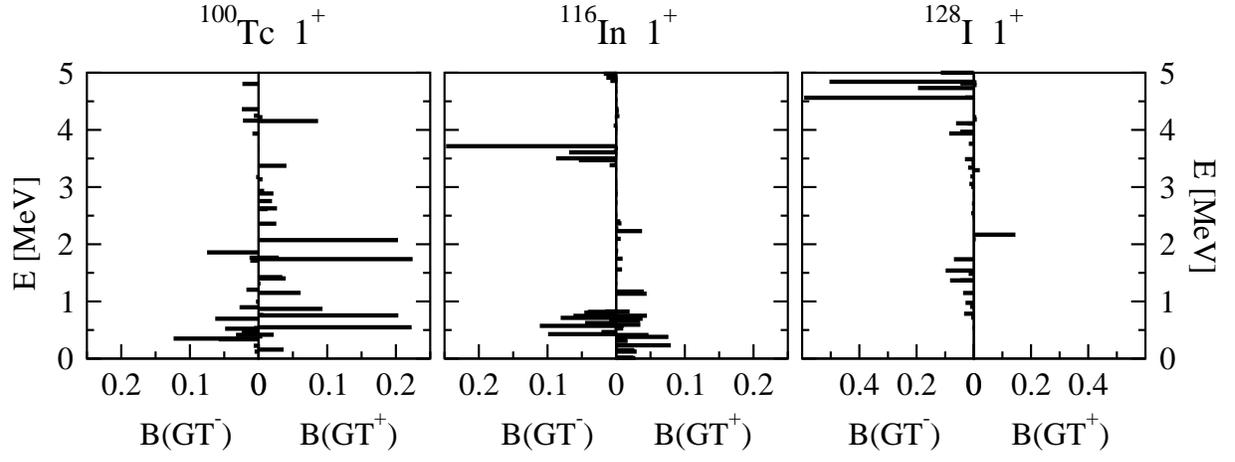}
\caption[]{$B(GT^{\pm})$ strength distributions in the intermediate
nucleus $^{100}$Tc (left), $^{116}$In (middle), and $^{128}$I (right)
of the double-beta decay $^{100}$Mo$\to^{100}$Ru,
$^{116}$Cd$\to^{116}$Sn, and $^{128}$Te$\to^{128}$Xe,
respectively.}
\label{espectrosGT}
\end{figure*}

A graphical representation of the calculations involved in Eq. (\ref{mgt}) 
is shown in Fig.~\ref{espectrosGT}, featuring the theoretical $1^+$ 
spectrum of the intermediate nucleus $^{100}$Tc in the decay 
$^{100}$Mo$\to ^{100}$Ru in the left panel, the intermediate nucleus 
$^{116}$In in the decay $^{116}$Cd$\to ^{116}$Sn in the middle panel, and
the intermediate nucleus $^{128}$I in the decay $^{128}$Te$\to ^{128}$Xe 
in the right panel. The vertical scale is the excitation energy of the 
intermediate nucleus, the horizontal scale represents the $B(GT)$ strength 
of the $1^+$ QRPA states of the intermediate nucleus as obtained from the 
$GT^-$ transitions in the initial double-beta emitter (shown from the 
central axis to the left), and from the $GT^+$ transitions in the final 
double-beta partner (shown from the central axis to the right).
The larger this strength, the more relevant the corresponding contribution 
to the total double-beta decay matrix element, provided there are states 
coming from the other branch of the decay with similar wave functions 
(giving rise to non-negligible overlaps) and carrying non-negligible 
strengths.

\subsection{Single electron energy distributions}

In Ref.~\cite{simk_ssd}, the SSD hypothesis was tested using an exact 
treatment of the energy denominator in the nuclear matrix element,
instead of the usual treatment of approximating the lepton energy by 
an average energy. It was shown that using the exact treatment of the energy 
denominator, the $2\nu\beta\beta$ decay half-life in the case of $^{100}$Mo,
is reduced by a factor of 20\% for ground state to ground state decay.
The single electron energy distributions of the outgoing electrons 
were also analyzed, and the predictions of different assumptions,
SSD and higher state dominance (HSD)\cite{simk_ssd}, were compared.
In this work we extend this analysis to the case of pnQRPA 
theoretical calculations of the $2\nu\beta\beta$ decay rates with 
a correct treatment of the energy denominators and compare these 
predictions with those of SSD and HSD.

The formalism needed to calculate the differential $2\nu\beta\beta$
decay rates was developed in Ref.~\cite{simk_ssd}.
Here we only write the basic expressions used for the evaluation
of the normalized decay rates.

The differential $2\nu\beta\beta$ decay rate to $0^+$ ground state 
can be written as 

\begin{eqnarray}
dW = a_{2\nu} F(Z_f,E_{e1}) F(Z_f,E_{e2})~{\cal M}~d\Omega,   
\end{eqnarray}
where $a_{2\nu}=(G_\beta g_A)^4 m_e^9 /(64 \pi^7)$ 
and $G_\beta=G_F \cos\theta_c$ ($G_F$ is Fermi constant,
$\theta_c$ is Cabbibo angle). $F(Z_f,E_e)$ denotes the 
relativistic Coulomb factor. The phase space factor is given by  
\begin{eqnarray}
d\Omega &=& \frac{2}{m^{11}_e}
E_{e1} p_{e1}~ E_{e2} p_{e2}~ E^2_{\nu 1}~ E^2_{\nu 2}
~\delta (E_{e1} + E_{e2} + E_{\nu 1} + E_{\nu 2} + E_{f} - E_{i})
\times \nonumber \\
&& ~~~~~~~~~~d E_{e1}~d E_{e2}~d E_{\nu 1}~d E_{\nu 2}.
\end{eqnarray}
${\cal M}$ consists of the products of nuclear matrix elements: 
\begin{eqnarray}
{\cal M} &=&
\frac{m^2_e}{4} \left[ |{\cal K}+{\cal L}|^2 
+ \frac{1}{3}|{\cal K}-{\cal L}|^2 
\right] \, ,
\end{eqnarray}
where ${\cal K} ({\cal L})$ denotes nuclear matrix elements with an energy
denominator given by $K_m$ ($L_m$),

\begin{eqnarray}
K_m \equiv [E_m  - E_i + E_{e1} + E_{\nu 1}]^{-1} + [E_m  - E_i + E_{e2} + 
E_{\nu 2}]^{-1},
\nonumber\\
L_m \equiv [E_m  - E_i + E_{e2} + E_{\nu 1}]^{-1} + [E_m  - E_i + E_{e1} + 
E_{\nu2}]^{-1}. 
\label{propf}
\end{eqnarray}

The normalized total decay rate can be written as
\begin{equation}
P = \frac{1}{W}\frac{dW}{dE_e} \, ,
\end{equation}
where the full decay probability $W$ is given by 
\begin{eqnarray}
W  = \frac{2 a_{2\nu}}{m^{11}_e}
\int_{m_e}^{E_i-E_f-m_e} 
F(Z_f,E_{e1}) p_{e1} E_{e1} dE_{e1} \times \nonumber\\ 
\int_{m_e}^{E_i-E_f-E_{e1}} F(Z_f,p_{e2}) p_{e2} E_{e2} dE_{e2}
\int_{0}^{E_i-E_f-E_{e1}-E_{e2}} {\cal M}~ E_{\nu 2}^2  E_{\nu 1}^2  
d E_{\nu 1} \, .
\end{eqnarray}

\begin{figure*}
\centering \includegraphics[width=140mm]{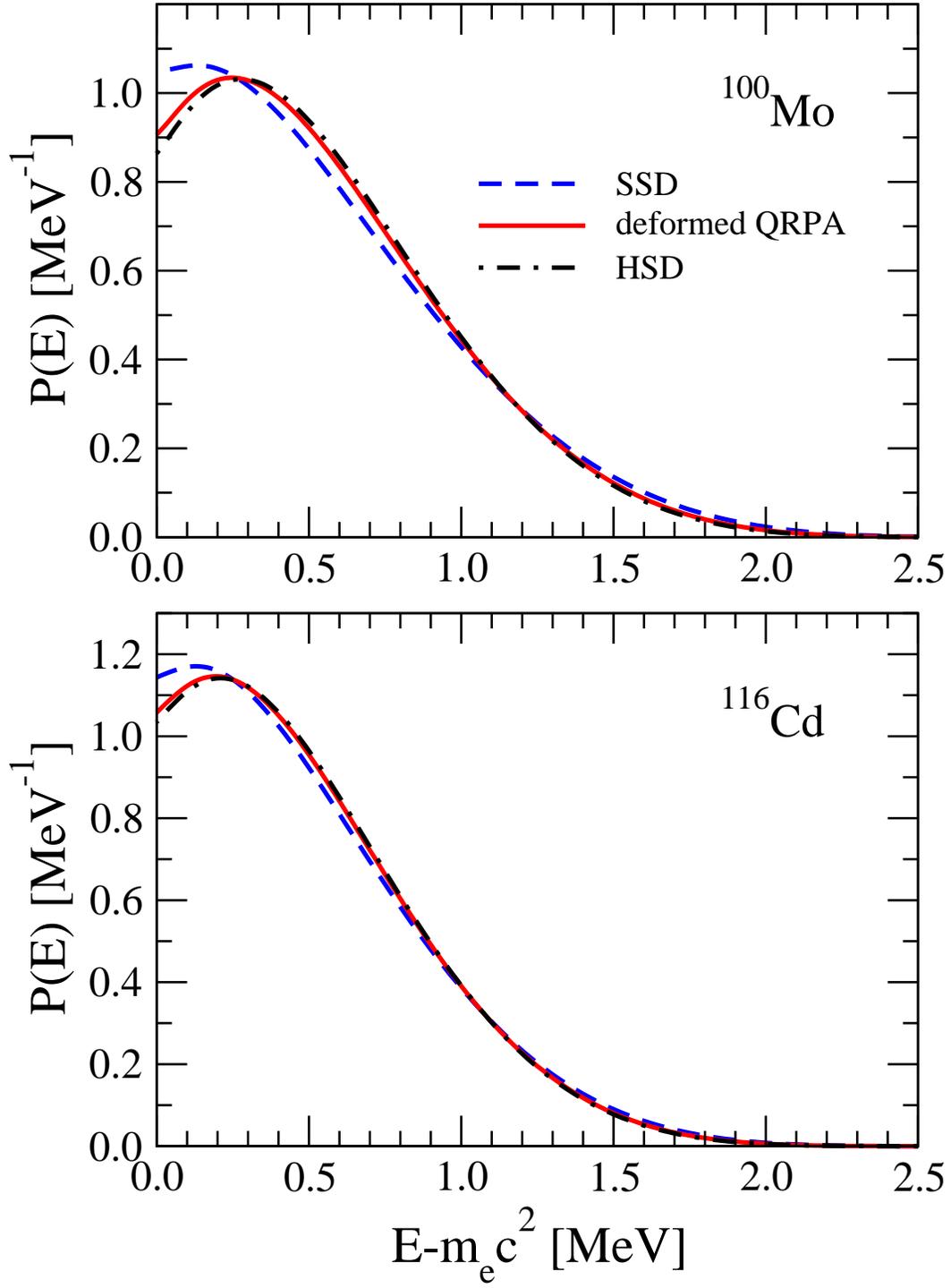}
\caption[]{Single electron differential decay rate normalized to the 
total decay rate for the $2\nu\beta\beta$ decay to the $0^+$
ground state in $^{100}$Mo$\to^{100}$Ru  (upper panel) and
$^{116}$Cd$\to^{116}$Sn (lower panel).}
\label{lepton}
\end{figure*}

In Fig.~\ref{lepton} we present the single electron spectrum of
the emitted electrons calculated within SSD, HSD, and total pnQRPA
for the $2\nu\beta\beta$ of $^{100}$Mo (upper panel) and 
$^{116}$Cd (lower panel).
The different behavior observed at small electron energies for
different assumptions is of the order of few percent. This 
accuracy is now accessible in the NEMO-3 experiment \cite{nemo}, 
where a large amount of events for the $2\nu\beta\beta$ decay of
$^{100}$Mo has been collected.

\section{Test of LLSD in double-beta emitters}

As an extension of the analysis of the SSD hypothesis, we have also
studied the contributions to the $2\nu\beta\beta$ decay matrix element
through the low-lying intermediate states for all confirmed 
$2\nu\beta\beta$ partners, also in the cases in which 
the ground state of the corresponding 
intermediate nucleus is not a $1^+$ state. To this end, we plot in 
Fig.~\ref{bb_accum} for all the nuclei under study 
the $2\nu\beta\beta$ decay matrix elements as a function of the 
excitation energy of the intermediate nucleus taken into account in the
calculation. As in the previous section, a deformed Skyrme-SLy4 HF mean field 
has been used to describe the initial and final ground states, with pairing
correlations treated in BCS approximation. 
In general we use the ground states for initial and final nuclei with a
shape that minimize the HF+BCS energy. The corresponding results are shown
by the solid lines. In some instances like $^{48}$Ti and $^{76}$Se, the 
minima correspond to a spherical shape ($\beta =0$), while experimental data
 \cite{rag89} indicate that these nuclei have a non negligible deformation
in the ground state ($\beta \sim 0.15$). In these cases we also show results
obtained with the pnQRPA calculations based on the HF+BCS solutions
corresponding to these non-zero deformations ($\beta =0.17$ for $^{48}$Ti
and $\beta =0.14$ for $^{76}$Se). The corresponding results shown by dashed
lines in Fig.~\ref{bb_accum} are seen to agree much better with experiment.
It is important to remark that particularly in the case of  $^{48}$Ti,
the HF+BCS energy has a rather shallow minimum in the deformation range from
$\beta=0$ to $\beta=0.17$, so that both solutions are equally plausible
ground states of  $^{48}$Ti.
The grey area in the plots indicates the experimental range of the
$2\nu\beta\beta$ decay matrix element. The two limiting horizontal lines
have been deduced from
experimental half-lives \cite{bar0206} using two different values of
the constant $g_A$, namely $g_A$=1.25 (bare value) and $g_A$=1.00
(quenched value). In this case the theoretical results are not quenched.

Typically, the $2\nu\beta\beta$ decay matrix elements shown in the plots
increase with the excitation energy of the intermediate states until
they reach a constant value right after a rather prominent final step located
between 15 and 20 MeV. This last step, especially important for the
double-beta partners with $A$ = 128, 130 and 136, is related to the
$GT^-$ giant resonance. The final value of the $2\nu\beta\beta$ decay 
matrix element lies within
the experimental region in most cases. Important deviations are only found
for the nuclei with $A$ = 130 and 136. In the cases of $A$ = 48 and $A$ = 76,
as already mentioned, the prolate shapes of the daughter nuclei give a 
better result than the
spherical shapes. Concerning the low energy contributions to
the matrix element, the fastest increase appears in the double-beta
partners with $A$ = 48, 96, 100, and 116, where around a 
60$\%$ of the total matrix element is reached within an excitation energy 
range of 2 MeV, and to a lesser extent the $A$=150 case.  In the other
double-beta partners the contributions to the matrix element are more
spread and the increase is slower.
In general, important contributions appear from relatively
high energies around the position of the GT resonance. 
Thus, although neither SSD nor LLSD hypotheses are clearly fulfilled in
our theoretical results, one can see a tendency in most cases to exhaust
at least 50\% of the total  $2\nu\beta\beta$ matrix elements in a low excitation
energy region of 5 MeV. The only salient exceptions to this rule
are the cases of $A$=82, 130, and 136.

\section{conclusions}

In this work we have studied the $2\nu\beta\beta$ decay matrix elements
within a selfconsistent Skyrme Hartree-Fock calculation with pairing and
deformation, and with residual $ph$ and $pp$ interactions treated in pnQRPA.
The analysis has been focused on the study of the validity of SSD and LLSD.
While the former has been tested on the $2\nu\beta\beta$ decay of $^{100}$Mo,
$^{116}$Cd, and $^{128}$Te, the latter has been checked on all the measured 
$2\nu\beta\beta$ emitters. Confirmation of that hypothesis has important 
consequences both experimentally and theoretically, since it could help to 
drastically simplify the experimental effort to get good estimates of 
$2\nu\beta\beta$ half-lives on one hand, and to simplify the theoretical 
description of the intermediate nucleus on the other.

With this aim, we have studied first the $B(GT^\pm)$ strength distributions 
of the intermediate $2\nu\beta\beta$ nuclei $^{100}$Tc, $^{116}$In, and 
$^{128}$I in the low range of excitation energies, comparing the theoretical 
results with the available experimental information from the direct decay 
and from charge-exchange reactions. The calculations reproduce fairly well 
the experimental distributions at low energy as well as the position and 
total strength of the GT resonances.

The result of the analysis performed concludes that SSD is experimentally
realized in the case of $^{100}$Mo because its experimental $2\nu\beta\beta$ 
half-life is roughly accounted for by the measured $B(GT)$ strengths
connecting the ground states of the nuclei involved. On the contrary, 
the SSD half-lives are underestimated in the case of $^{116}$Cd and are
overestimated in the case of $^{128}$Te. However, these
results are still not conclusive since the uncertainties arising from the
insufficient precision of the experimental measurements of ft-values, of 
$2\nu\beta\beta$ half-lives, and especially because of the uncertainties of 
the $B(GT)$ strength extracted from charge-exchange reactions.

From the theoretical side, 
the conclusion of our results is that clear evidences for SSD and LLSD are
not found within the present approach. This is at variance with the conclusion
in Ref. \cite{civi_suho_98}, where strong support of SSD was claimed. However,
we would like to point out that our theoretical LLSD $2\nu\beta\beta$ matrix 
elements agree up to factors of the order of 2 with our total matrix elements. 
This finding is in qualitative agreement with the results obtained in previous 
studies \cite{civi_suho_98,simk_ssd}. Furthermore, in Ref. \cite{civi_suho_98} 
it is stated that SSD results agree quite well with the total matrix elements
even though they find in some cases discrepancies up to a factor of 2.

With only two exceptions ($A$=130, 136), our calculations reproduce the 
experimental $2\nu\beta\beta$ half-lives when the whole energy range of 
excitation energies in the intermediate nucleus is considered. 
Within the considered nuclear model we find that important contributions 
to the $2\nu\beta\beta$ decay matrix elements arise from relatively high 
excitation energies in several cases, not supporting dominance of single 
or even low-lying states. Further progress in both experimental and 
theoretical sides is still needed to clarify the importance of the
contributions of the intermediate nucleus states to the $2\nu\beta\beta$ 
decay matrix elements. 
In the experimental case, this progress can be achieved by completing 
the experiments with high resolution
charge-exchange reactions to more nuclei of interest and extending the 
measurements of GT strength distributions up to the resonance regions. 
For further progress in the field, measurements of the occupation numbers 
in particle-transfer reactions \cite{occu} are also of great importance.
In the theoretical case improvements are needed in the treatment of the 
residual forces by using realistic interactions and improved
parametrizations, especially in the particle-particle sector.
The role of the axial-vector coupling strength should be also studied
further because systematic indications in favor of strong quenching have 
been recently reported \cite{quenching}.

The single electron energy distributions of the outgoing electrons in
the $2\nu\beta\beta$ decay processes have been investigated under 
various assumptions (SSD, HSD), as well as from theoretical calculations
using an exact treatment of the energy denominators of the perturbation
theory, i.e., without factorization of the nuclear structure and phase 
space integration calculations.
A more comprehensive analysis of NEMO-3 data including the theoretical 
predictions presented in this paper could help
to confirm or rule out the possible realizations of SSD and LLSD.

\begin{acknowledgments}
This work was supported by Ministerio de Ciencia e Innovaci\'on
(Spain) under Contract No. FIS2005-00640. It was also supported in part 
by the EU ILIAS project under contract RII3-CT-2004-506222.
O.M. and R.A.R. thank Ministerio 
de Ciencia e Innovaci\'on (Spain) for financial support. 
\end{acknowledgments}

\newpage

\begin{table}[t]
\caption{Experimental Gamow-Teller matrix elements $M(GT^-)$ and $M(GT^+)$ 
for the transitions connecting the $1^+$ ground state in $^{100}$Tc with
the $0^+$ ground states in $^{100}$Mo and $^{100}$Ru. We also show the 
matrix elements
$M^{2\nu\beta\beta}_{GT}$ [MeV$^{-1}$] and the corresponding half-lives 
$T_{1/2}^{2\nu\beta\beta}$ [y] obtained with aproximated (SSD1) and exact
(SSD2) energy denominators. The experimental half-life is 
$T_{1/2}^{2\nu\beta\beta}=7.1\times 10^{18}$ y \cite{bar0206}, 
and its corresponding matrix element is 
$(M^{2\nu\beta\beta}_{GT})_{\rm exp}= 0.241$ MeV$^{-1}$.
While the half-lives are independent on $g_A$, the matrix elements have 
been obtained using  $g_A$=1.25.}
\label{SSD100}
\begin{tabular}{ccccc}\cr
$M(GT^-)$ & $M(GT^+)$ & $M^{2\nu\beta\beta}_{GT}$ &
 $T_{1/2}^{2\nu\beta\beta}$ (SSD1) & $T_{1/2}^{2\nu\beta\beta}$ (SSD2) \cr
\hline
\cr
0.57\footnote{from ($^3$He,$t$) \cite{aki97}}  & 
0.55\footnote{from $\log ft_{\beta ^-}$ \cite{sin08}}& 0.19 & 
1.1 $\times  10^{19}$ & 8.8 $\times  10^{18}$  \cr
\cr
0.65\footnote{from $\log ft_{\rm EC}$ \cite{garcia}} & 0.55\footnotemark[2] 
& 0.21 &   9.3 $\times  10^{18}$ & 7.2 $\times  10^{18}$\cr
\hline
\end{tabular}
\end{table}

\begin{table}[t]
\caption{Same as Table \ref{SSD100}, but for $A=116$. In this case
the $2\nu\beta\beta$ half-life is 
$T_{1/2}^{2\nu\beta\beta}=3\times 10^{19}$ y  \cite{bar0206}, 
and its corresponding matrix element is 
$(M^{2\nu\beta\beta}_{GT})_{\rm exp}= 0.127$ MeV$^{-1}$.}
\label{SSD116}
\begin{tabular}{ccccc}\cr
$M(GT^-)$ & $M(GT^+)$ & $M^{2\nu\beta\beta}_{GT}$ 
&  $T_{1/2}^{2\nu\beta\beta}$ (SSD1) & $T_{1/2}^{2\nu\beta\beta}$ (SSD2) \cr
\hline
\cr
 0.18\footnote{from ($^3$He,$t$) \cite{aki97}}  &   
0.51\footnote{from $\log ft_{\beta ^-}$ \cite{bla01}, used by \cite{rak05}} & 
0.05 &   1.9 $\times  10^{20}$ &   1.7 $\times  10^{20}$ \cr
\cr
 0.51\footnote{from $(p,n)$ \cite{sas07}}  &   0.51\footnotemark[2]  & 
0.14 &  2.4 $\times  10^{19}$ &  2.1 $\times  10^{19}$ \cr
\cr
 0.69\footnote{from $\log ft_{\rm EC}$ \cite{bhatta}}  &  0.51\footnotemark[2]  
& 0.19 &  1.3 $\times  10^{19}$   &  1.2 $\times  10^{19}$  \cr
\hline
\end{tabular}
\end{table}

\begin{table}[t]
\caption{Same as Table \ref{SSD100}, but for $A=128$. In this case
the $2\nu\beta\beta$ half-life is 
$T_{1/2}^{2\nu\beta\beta}=2.5\times 10^{24}$ y  \cite{bar0206}, 
and its corresponding matrix element is 
$(M^{2\nu\beta\beta}_{GT})_{\rm exp}= 0.043$ MeV$^{-1}$.}
\label{SSD128}
\begin{tabular}{ccccc}\cr
$M(GT^-)$ & $M(GT^+)$ & $M^{2\nu\beta\beta}_{GT}$ 
&  $T_{1/2}^{2\nu\beta\beta}$ (SSD1) & $T_{1/2}^{2\nu\beta\beta}$ (SSD2) \cr
\hline
\cr
 0.41\footnote{from ($p,n$) \cite{madey}}  &   
0.10\footnote{from $\log ft_{\beta ^-}$ \cite{kanbe01}} & 
0.024 &  8.1 $\times  10^{24}$  &  7.8 $\times  10^{24}$ \cr
\cr
 0.33\footnote{from $\log ft_{\rm EC}$ \cite{kanbe01}}  &  0.10\footnotemark[2]  
& 0.019 &  1.3 $\times  10^{25}$   &  1.2 $\times  10^{25}$ \cr
\hline
\end{tabular}
\end{table}

\begin{table}[t]
\caption{Theoretical (quenched) and experimental accumulated single 
Gamow-Teller strengths $B(GT^-$) and $B(GT^+$) up to the different measured 
excitation energies of the intermediate nucleus. The corresponding values 
of the double GT matrix element $M^{2\nu\beta\beta}_{GT}$  [MeV$^{-1}$]
up to the same excitation energies for $^{100}$Mo $\to ^{100}$Ru are 
also given.}
\label{LLSD100}
\begin{tabular}{ccccccc} \cr
& \multicolumn{2}{c} { } & \multicolumn{2}{c} { } & \multicolumn{2}{c} {}
\cr
& \multicolumn{2}{c} {$\sum$ B(GT$^-$)} & \multicolumn{2}{c} {$\sum$ 
B(GT$^+$)} & \multicolumn{2}{c} {$M^{2\nu\beta\beta}_{GT}$ }  \cr
\cline{2-3} \cline{4-5} \cline{6-7} \cr
E [MeV] &  th.  &  exp.\footnote{from ($^3$He,t) \cite{aki97}} &  th. & 
exp. & th. & exp.   \cr
\hline
\cr
1.4 & 0.42 &  0.46  & 0.73 &  -  & 0.12 & - \cr
\cr
\; 2.6 \; & \; 0.53 \; & \;  0.69\;   &\;  1.26\;  & \;  - \;  & \; 0.15\; 
&\;  - \;   \cr
\hline
\end{tabular}
\end{table}

\begin{table}[t]
\caption{Same as in Table \ref{LLSD100}, but for $A=116$.}
\label{LLSD116}
\begin{tabular}{ccccccc} \cr
& \multicolumn{2}{c} { } & \multicolumn{2}{c} { } & \multicolumn{2}{c} {}
\cr
& \multicolumn{2}{c} {$\sum$ B(GT$^-$)} & \multicolumn{2}{c} {$\sum$ 
B(GT$^+$)} & \multicolumn{2}{c} {$M^{2\nu\beta\beta}_{GT}$}  \cr
\cline{2-3} \cline{4-5} \cline{6-7} \cr
E [MeV]
& th. & exp.\footnote{from ($^3$He,t) \cite{aki97}} & th. & 
exp.\footnote{from ($d,^2$He) \cite{rak05}}
& th. & exp. \cr
\hline
\cr
0.7 & 0.29 &  0.03   & 0.50 & 0.33 
 & 0.03 & 0.05    \cr
1.0 & 0.56 &  0.15  & 0.56 & 0.44 & 0.06 & 0.09    \cr
2.2 & 0.56 &  0.32  & 0.69 & $<$0.51 & 0.06 & $<$0.12    \cr
\; 3.0 \; & \; 0.56\;  & \;  0.32\;   & \;0.75\;  & \; $<$0.72 \; 
& \; 0.06 \; & \; $<$0.12 \;   \cr
\hline
\end{tabular}
\end{table}

\begin{table}[t]
\caption{Same as in Table \ref{LLSD100}, but for $A=128$.}
\label{LLSD128}
\begin{tabular}{ccccccc} \cr
& \multicolumn{2}{c} { } & \multicolumn{2}{c} { } & \multicolumn{2}{c} {}
\cr
& \multicolumn{2}{c} {$\sum$ B(GT$^-$)} & \multicolumn{2}{c} {$\sum$ 
B(GT$^+$)} & \multicolumn{2}{c} {$M^{2\nu\beta\beta}_{GT}$}  \cr
\cline{2-3} \cline{4-5} \cline{6-7} \cr
E [MeV]
& th. & exp.\footnote{from ($p,n$) \cite{madey}} & th. & 
exp. & th. & exp. \cr
\hline
\cr
0.58 & 0.004 &  0.22   & 0.014 & -  & 0.000  & -    \cr
1.09 & 0.094 &  0.52   & 0.021 & -  & 0.004  & -    \cr
1.61 & 0.382 &  0.64   & 0.027 & -  & 0.012  & -    \cr
\; 2.56 \; & \; 0.468 \;  & \;  1.07 \;   & \; 0.187 \;  & \; - \; 
& \; 0.021 \; & \; - \;   \cr
\hline
\end{tabular}
\end{table}

\begin{thebibliography}{00}

\bibitem{osci}
Fukuda Y {\em et al} 1998 {\em Phys. Rev. Lett.} {\bf 81} 1562 \\ 
Ahn M H {\em et al} 2006 {\em Phys. Rev.} D {\bf 74} 072003 \\ 
Ahmad Q R {\em et al} 2002 {\em Phys. Rev. Lett.} {\bf 89} 011301 \\
Ahmad Q R {\em et al} 2002 {\em Phys. Rev. Lett.} {\bf 89} 011302 \\ 
Araki T {\em et al} 2005 {\em Phys. Rev. Lett.} {\bf 94} 081801

\bibitem{2bb1}
Suhonen J and Civitarese O 1998 {\em Phys. Rep.} {\bf 300} 123
\bibitem{2bb2}
Faessler A and \v Simkovic F 1998 {\em J. Phys. G: Nucl. Part. Phys.} {\bf 24} 2139
\bibitem{2bb3}
Elliott S and Vogel P 2002 {\em Ann. Rev. Nucl. Part. Phys.} {\bf 52} 115 

\bibitem{sm1}
Haxton W C and Stephenson Jr. G J 1984  {\em Prog. Part. Nucl. Phys.} {\bf 12} 409
\bibitem{sm2}
Caurier E, Nowacki F, Poves A and Retamosa J 1996 {\em Phys. Rev. Lett.} {\bf 77} 1954 

\bibitem{pnqrpa1}
Halbeib J A and Sorensen R A 1967 {\em Nucl. Phys.} A {\bf 98} 542
\bibitem{pnqrpa2}
Engel J, Vogel P and Zirnbauer M R 1988 {\em Phys. Rev.} C {\bf 37} 731
\bibitem{pnqrpa3}
Muto K and Klapdor H V 1988 {\em  Phys. Lett.} B {\bf 201} 420
\bibitem{pnqrpa4}
Suhonen J, Taigel T and Faessler A 1988 {\em  Nucl. Phys.} A {\bf 486} 91

\bibitem{aba84} Abad J, Morales A, Nu\~nez-Lagos R, Pacheco A F 1984
{\em An. Fis.} A {\bf 80} 9

\bibitem{garcia}
Garc\'{\i}a A {\em et al} 1993 {\em  Phys. Rev.} C {\bf 47} 2910

\bibitem{bhatta}
Bhattacharya M {\em et al} 1998 {\em  Phys. Rev.} C {\bf 58} 1247

\bibitem{kanbe01} Kanbe M and Kitao K 2001 {\em  Nucl. Data Sheets} {\bf 94}
227

\bibitem{civi_suho_98}
Civitarese O and Suhonen J 1998 {\em Phys. Rev.} C {\bf 58} 1535\\
Civitarese O and Suhonen J 1999 {\em Nucl. Phys.} A {\bf 653} 321

\bibitem{simk_ssd}
Domin P, Kovalenko S, \v Simkovic F and Semenov S V 2005 {\em Nucl. Phys.} A 
{\bf 753} 337 \\
\v Simkovic F, Domin P and Semenov S V 2001 {\em J. Phys. G: Nucl. Part. Phys.} 
{\bf 27} 2233

\bibitem{frekers} Rakers S {\em et al}  2004 {\em Phys. Rev.} C {\bf 70} 054302 \\
Grewe E W {\em et al} 2006 {\em Prog. Part. Nucl. Phys.} {\bf 57} 260\\
Grewe E W {\em et al} 2007 {\em Phys. Rev.} C {\bf 76} 054307 \\
Grewe E W {\em et al} 2008 {\em Phys. Rev.} C {\bf 77} 064303

\bibitem{madey} Madey R {\em et al} 1989 {\em  Phys. Rev.} C {\bf 40} 540

\bibitem{aki97} Akimune H, Ejiri H, Fujiwara M, Daito I,
Inomata T, Hazama R, Tamii A, Toyokawa H and Yosoi M 1997 {\em Phys. Lett.} B
{\bf 394} 23

\bibitem{sas07} Sasano M{\em et al} 2007 {\em Nucl. Phys.} A {\bf 788} 76c

\bibitem{rak05} Rakers S {\em et al} 2005 {\em Phys. Rev.} C {\bf 71} 054313

\bibitem{vautherin} Vautherin D and Brink D M 1972 {\em Phys. Rev.} C 
{\bf 5} 626\\
 Vautherin D 1973 {\em Phys. Rev.} C {\bf 7} 296

\bibitem{single}  Sarriguren P, Moya de Guerra E, Escuderos A and 
Carrizo A C 1998 {\em  Nucl. Phys.} A {\bf 635} 55\\ 
Sarriguren P, Moya de Guerra E and Escuderos A 1999 {\em Nucl. Phys.} A {\bf A58} 13 \\
Sarriguren P, Moya de Guerra E and Escuderos A 2001 {\em Nucl. Phys.} A {\bf A691} 631\\
Sarriguren P, Moya de Guerra E and Escuderos A 2001 {\em Phys. Rev.} C {\bf 64} 064306 

\bibitem{simk_def} \v Simkovic F, Pacearescu L and Faessler A 2004 {\em Nucl. Phys.} 
A {\bf 733} 321

\bibitem{nakada} Nakada H, Sebe T and Muto K 1996 {\em Nucl. Phys.} A {\bf 607} 235

\bibitem{raquel} \'{A}lvarez-Rodr\'{i}guez R, Sarriguren P, Moya
de Guerra E, Pacearescu L, Faessler A and \v Simkovic F 2004 {\em Phys. Rev.} 
C {\bf 70} 064309

\bibitem{bmot} Bohr A and Mottelson B  1975 {\em Nuclear Structure} (Benjamin,
New York)

\bibitem{emoya} Moya de Guerra E 1986 {\em Phys. Rep.} {\bf 138} 293 \\
Villars F 1966 {\em Many-body description of nuclear structure and reactions},
ed. C. Bloch (Academic Press, N.Y.)

\bibitem{berdi} Moya de Guerra E and Kowalski S 1979 {\em Phys. Rev.} C {\bf 20}
357 \\
Berdichevsky D, Sarriguren P, Moya de Guerra E, Nishimura M,
and Sprung D W L 1988 {\em Phys. Rev.} C {\bf 38} 338\\
Sarriguren P, Graca E, Sprung D W L, Moya de Guerra E and Berdichevsky D 1989
{\em Phys. Rev.} C {\bf 40} 1414 

\bibitem{cha98} Chabanat E, Bonche P, Haensel P, Meyer J and  
Schaeffer R 1998 {\em Nucl. Phys.} A {\bf 635} 231

\bibitem{au03} Audi G, Bersillon O, Blachot J and Wapstra A H 2003
{\em Nucl. Phys.} A {\bf 729} 3

\bibitem{sin08} Singh B 2008 {\em Nucl. Data Sheets} {\bf 109} 297

\bibitem{bar0206} Barabash A S 2002 {\em Czech. J. Phys.} {\bf 52} 567\\
Barabash A S 2006 arXiv:nuclex/0602009v2

\bibitem{bla01} Blachot J 2001 {\em  Nucl. Data Sheets} {\bf 92} 455

\bibitem{nemo} Arnold R {\em et al} 2005 {\em Phys. Rev. Lett.} {\bf 95}
182302 \\
Arnold R {\em et al} 2007 {\em Nucl. Phys.} A {\bf 781} 209 

\bibitem{rag89} Raghavan P 1989 {\em Atomic and Nuclear Data Tables} {\bf 42}
  189 \\
Stone N J 2001 {\em Table of Nuclear Moments}
  www.nndc.bnl.gov/nndc/stone$\_$moments

\bibitem{occu}  Freeman  S J {\em et al} 2007 {\em  Phys. Rev.} C {\bf 75} 
051301(R) \\
Schiffer J P {\em et al} 2008 {\em Phys. Rev. Lett.} {\bf 100} 112501

\bibitem{quenching} Faessler A, Fogli G L, Lisi E, Rodin V, 
Rotunno A M and \v Simkovic F 2008 {\em J. Phys. G: Nucl. Part. Phys.} 
{\bf 35} 075104

\end{thebibliography}
\end{document}